\documentclass[12pt,a4paper]{article}
\usepackage{jheppub}

\usepackage{graphicx}
\usepackage{amsmath}
\usepackage{amssymb,amsbsy}
\usepackage{epsfig}
\usepackage{subcaption}
\usepackage{xcolor}
\usepackage{braket}
\usepackage{esint}
\usepackage{xcolor}
\usepackage{dsfont}

\newcommand{\dd}{\mathrm{d}}

\title{D-branes (or not) in the non-Abelian T-dual of the $SU(2)$ WZW model}
\author{Benjo Fraser}
\affiliation{ 
 CUniverse Group\\
 Department of Physics\\
  Faculty of Science\\
  Chulalongkorn University\\
  Bangkok 10330, Thailand
}
\emailAdd{benjojazz@gmail.com}
\abstract{Following on from arXiv:1805.03657, we consider open strings in the non-Abelian T-dual of the $SU(2)_k$ WZW model, with respect to the vector $SU(2)$ isometry. Since in this case the dual theory has an exact CFT description, we look at the chiral algebra-preserving D-branes. The general conclusion is that a large set of branes, while consistent boundary conditions on the worldsheet, are sent to infinity in the target space. This suggests inconsistency of the strict non-Abelian duality with open strings, in this case. }
\begin{document}
\maketitle
\section{Introduction}
WZW models and their cosets \cite{DiFrancesco:1997nk} are fruitful models to consider in string theory. This is because they enjoy an extension of the Virasoro algebra to larger chiral algebras, which respect to which they are rational. This allows the spectrum to be worked out exactly as a function of the level $k$, which typically parametrizes the curvature of the target manifold. It also provides, via arguments originally due to Cardy \cite{Cardy:1989ir}, an exact CFT description of certain open string sectors, i.e. D-brane configurations. D-branes possess charges conjectured \cite{Witten:1998cd} to be valued in the twisted K-theory of the target space, and in the case of the $SU(N)_k$ WZW model the `Cardy branes', with respect to appropriate subalgebras, were shown to furnish a complete set of representatives of this space of charges. 

Given this sort of solubility, is natural to explore these models as fully as possible. One avenue is Abelian and non-Abelian T-duality \cite{Giveon:1993ai}. It was shown in \cite{Sfetsos:1994vz}, that the classical action for the non-Abelian T-dual of the $SU(2)_k$ WZW model with respect to the vector $SU(2)$ isometry can be obtained as the limit of the coset model
\begin{align}\label{eq:coset}
\frac{\hat{\mathfrak{su}}(2)_k\oplus \hat{\mathfrak{su}}(2)_{\ell}}{\hat{\mathfrak{su}}(2)_{k+\ell}}
\end{align}
in which $\ell\rightarrow\infty$, and we must also zoom in close to the identity in the corresponding coset geometry, parametrizing the $SU(2)_{\ell}$ group element as $g_2=\mathds{1}+\mathrm{i}\, v/\ell$. The recent paper \cite{Fraser:2018cxn} developed a corresponding modular invariant truncation of the exact coset CFT. The torus partition (namely, the closed string sector) was calculated, and the fusion rules were discussed. 

The goal of the present paper is to consider open strings in the truncation, in the process fleshing out certain aspects of the picture. We fail to find a consistent picture unless the whole coset geometry is included. 
\section{D-branes in rational CFT}\label{sec:rational}
Cardy branes in rational CFT are constructed in the following way. We are interested in boundary states which preserve a given chiral algebra $\mathcal{A}$. In the closed string picture these correspond to boundary states $\ket{\mathcal{B}}$ satisfying
\begin{align}\label{eq:symmetricbc}
\left( J_n\, +\, \Omega (\bar{J}_{-n})\right)\ket{\mathcal{B}}\, =\, 0
\end{align}
where $J_n,\bar{J}_n$ are the (anti-)holomorphic modes of the algebra generators, and $\Omega$ is some outer automorphism. For us $\Omega$ will always be trivial. An orthogonal basis of solutions to \eqref{eq:symmetricbc} is given by the Ishibashi states $\ket{i}\rangle$, one for each representation $i$ of $\mathcal{A}$  appearing in the spectrum (we consider here the diagonal modular invariant). Consistency with modular transformations picks specific linear combinations of the $\ket{i}\rangle$ to be good boundary states for the CFT. These are the `Cardy states'\footnote{These also satisfy certain `sewing' constraints \cite{Gaberdiel:2002my}. }
\begin{align}\label{eq:cardy}
\ket{i}_C\, \equiv\, \frac{S_{ij}}{\sqrt{S_{0i}}}\ket{j}\rangle\quad .
\end{align}
The embedding of the D-brane in the target space may be deduced from identifying the zero-mode distributions of the Ishibashi states with the solutions of the scalar Laplacian - this yields functions which reduce at large level to delta functions supported on the D-brane worldvolume. 
\section{Branes in the WZW model}\label{sec:wzw}
\begin{figure}[hbtp]
\centering
\begin{subfigure}[tl]{0.4\textwidth}
\centering
\includegraphics[scale=0.7]{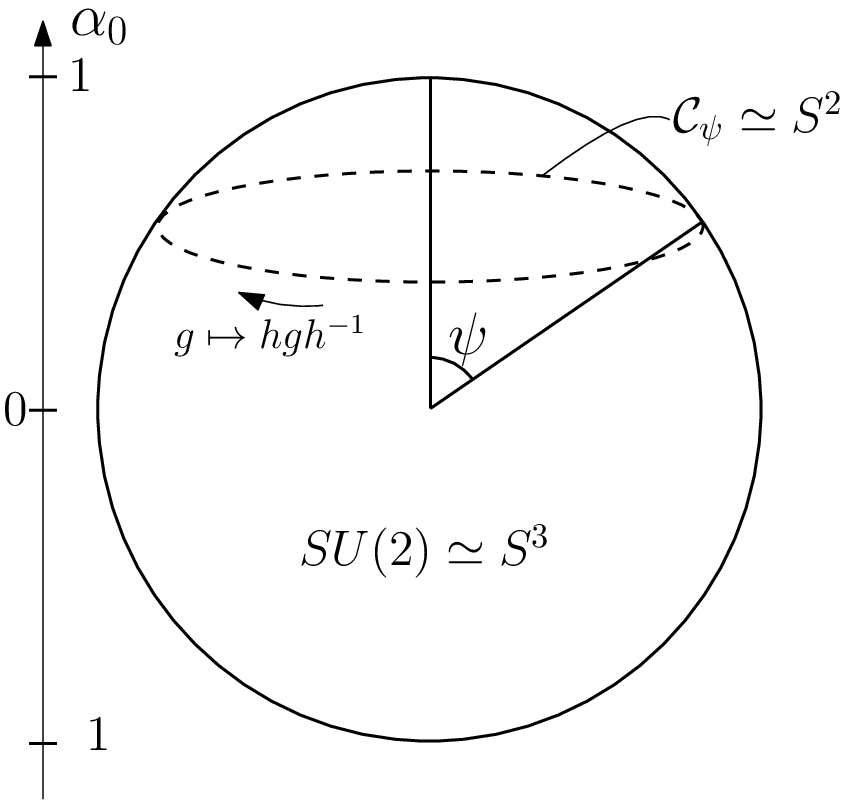}
\caption{A conjugacy class at polar angle $\psi=(\lambda/k)\pi$ on the $SU(2)$ group manifold.}
\label{fig:conjugacy}
\end{subfigure}
\hspace{0.5in}
\begin{subfigure}[tr]{0.4\textwidth}
\centering
\includegraphics[scale=0.7]{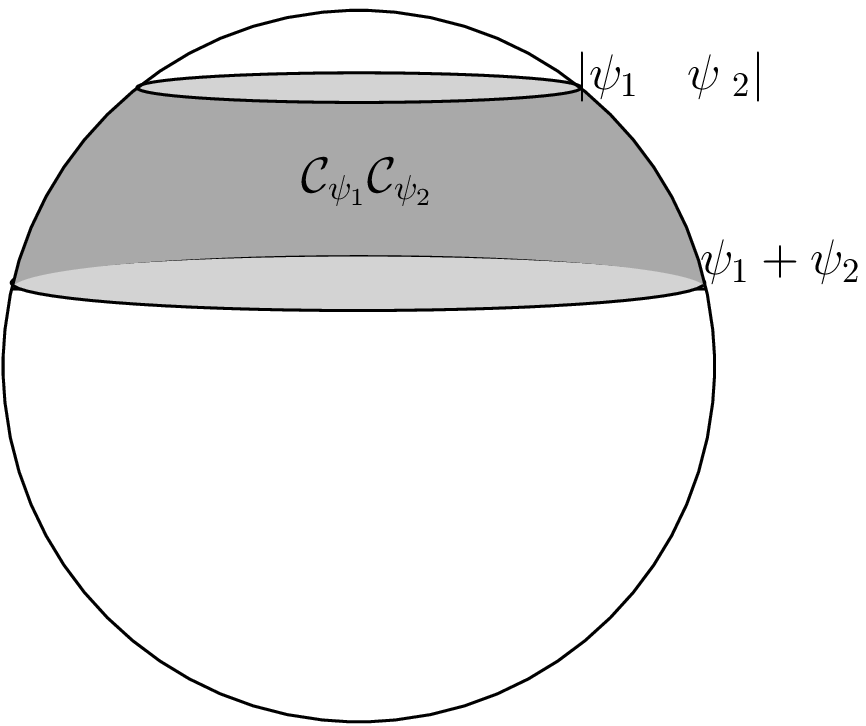}
\caption{A product of two conjugacy classes at polar angles $\psi_{1,2}$.}
\label{fig:conj_product}
\end{subfigure}
\caption{Conjugacy classes and products}
\end{figure}

The $SU(2)$ WZW model has target space $SU(2)\simeq S^3$, which we parametrize by embedding in $\mathbb{R}^4$ as $\alpha_0^2+|\pmb{\alpha}|^2=1$, and we define a polar angle $\alpha_0\equiv \cos\psi$. Representations of the chiral algebra $\hat{\mathfrak{su}}(2)_k$ are labelled by an integer $0\leq \lambda\leq k$, and the classical $\hat{\mathfrak{su}}(2)_k$-preserving Cardy brane worldvolumes are given by conjugacy classes $\mathcal{C}_{\lambda}=\mathcal{C}_{\psi=(\lambda/k)\pi}$ \cite{Alekseev:1998mc,Felder:1999ka,Stanciu:1999id}, which are the spaces of constant $\alpha_0$, each one isomorphic to $S^2$ except two points at each pole $\alpha_0=\pm 1$, as shown in fig \ref{fig:conjugacy}. 

Let us see this explicitly from the CFT construction. Here we use ideas from \cite{Felder:1999ka}: the basic point is to associate to each Ishibashi state a corresponding function on the target space geometry, and to use it in the sum \eqref{eq:cardy} to find the spatial distribution of the brane. On a group manifold $G$ these are identified by the Peter-Weyl theorem as the matrix elements of irreducible representations of $G$. The appropriate function corresponding to a representation of weight $\lambda$ is the group character in that representation, which should be normalized with respect to the Haar measure. We are interested in the classical limit ($k\rightarrow\infty$). Parametrizing the weights as $\lambda\equiv x k$, this gives an integral for the brane distribution:
\begin{align}
\mathcal{B}_{\delta}(\psi)\, =\, \frac{(2k^5)^{1/4}}{|SU(2)|}\int_0^1\dd x\sqrt{x}\,\frac{\sin(\pi k\lambda x)}{\sqrt{\sin(\pi x)}}\frac{\sin(k\psi x)}{\sin\psi}\quad .
\end{align}
At $k$ large this function localizes around the conjugacy class $\psi=(\lambda/k)\pi$, as found in \cite{Felder:1999ka}\footnote{There is a finite shift in the weight, but this becomes negligible for large $\lambda$, and we are not interested in these details.}

Let us now consider the large-$k$ theory where we truncate to primaries whose weight $\lambda$ scales like $k^{1/2}$, as discussed in \cite{Fraser:2018cxn}. The closed string states, because they have wavelength $\sim k^{-1/2}$ times the linear size ($\sim\sqrt{k}$) of the target manifold (in units of string length), are supported by a zoomed-in North pole geometry where we send
\begin{align}
\alpha_0^2&\equiv 1-\left(\frac{\rho}{\sqrt{k}}\right)^2&k&\rightarrow\infty\quad .
\end{align}
In this limit the geometry is simply flat $\mathbb{R}^3$ with radial coordinate $\rho$. The scaling of the weights means the corresponding branes have worldvolumes which are $S^2$s of finite radius centered around $\rho=0$. 
\section{D-branes in the coset}
We are interested in the coset \eqref{eq:coset}. The coset space is parametrized by two spheres, call them $S^3_k$ and $S^3_{\ell}$, divided out by the diagonal $SU(2)$ action
\begin{align}
(g_1,g_2)\sim (hg_1h^{-1},\, hg_2h^{-1})\quad .
\end{align}
We embed both of these spheres into $\mathbb{R}^4$ as $\alpha_0^2+|\pmb{\alpha}|^2=1$, $\beta_0^2+|\pmb{\beta}|^2=1$ respectively. Conjugation acts as a common rotation of the vectors $\pmb{\alpha}$ and $\pmb{\beta}$ while leaving $\alpha_0, \beta_0$ fixed. Modding out by the diagonal action means coordinates on the coset space are the three rotationally invariant combinations $\alpha_0, \beta_0, \gamma\equiv \pm\sqrt{|\pmb{\alpha\cdot\beta}|}$. The metric on the coset space was derived in \cite{Bardacki:1990wj} and is written in our coordinates in \cite{Polychronakos:2010fg}. It has no continuous isometries, but has an interchange symmetry between $S^3_k$ and $S^3_{\ell}$ when we also take $k\leftrightarrow\ell$. 

The coset chiral algebra $\mathcal{A}_c$ is derived from the group theoretic structure. Representations of $\mathcal{A}_c$ are labelled by three integers $\lambda,\mu,\nu$ labelling the $\hat{\mathfrak{su}}(2)_{k,\ell,k+\ell}$ quantum numbers respectively.  As discussed in \cite{Gawedzki:2001ye,Fredenhagen:2001kw,Wurtz:2005mt}, the classical Cardy branes are then products of conjugacy classes $(\mathcal{C}_{\lambda}\mathcal{C}_{\nu},\mathcal{C}_{\mu}\mathcal{C}_{\nu})$ (note that a conjugacy class is its own inverse). 

As mentioned in section \ref{sec:wzw}, a conjugacy class on one of the spheres, say $S^3_k$, is a space of constant $\alpha_0$. Multiplying two conjugacy classes produces a 3D annular region extending between two latitudes. The rule for these latitudes is similar to the addition of spins, but truncated at the North and South poles. Letting $\psi$ be the polar angle $\alpha_0\equiv \cos\psi$, we have
\begin{align}
\mathcal{C}_{\psi_1}\mathcal{C}_{\psi_2}\, : \, |\psi_1-\psi_2|\leq\psi\leq {\rm min}[\psi_1+\psi_2,2\pi-(\psi_1+\psi_2)]
\end{align}
which is depicted in figure \ref{fig:conj_product}. In particular, the square of a conjugacy class always covers the North pole. A classical coset Cardy brane therefore consists of a pair of such intervals on $S^3_k$ and $S^3_{\ell}$. 

In non-Abelian T-duality we are interested in the following limit of the quantum numbers \cite{Fraser:2018cxn}
\begin{align}\label{eq:quantumnumbers}
\mu&=\frac{\delta}{k}\ell & \nu-\mu &= n = \mathrm{finite} & \ell\rightarrow\infty\quad .
\end{align}
where $0\leq\delta\leq k$. In this limit the radius of $S^3_{\ell}$ is large. Because the difference between the two weights $\mu$ and $\nu$ is finite, the Cardy branes for these weights are North polar caps on $S^3_{\ell}$, apart from a small hole right at the pole. The caps extends down to a polar angle $\frac{\mu+\nu}{\ell}\pi\sim \frac{\delta}{k}\pi$ until $\delta=k/2$ when it covers the whole sphere. For higher $\delta$ it starts to shrink again, ending up as a point at the North pole for $\delta=k$. 

We now want to see these branes emerging from the CFT description. This has been considered for a general coset in \cite{Fuchs:2005gp}. Since we are looking at the lowest mode of the string, the appropriate functions to represent the Ishibashi states should be eigenfunctions of the scalar Laplacian on the coset geometry. These were found in \cite{Sfetsos:1994vz}.  Since we are mostly interested in conceptual points, we will only consider the simplest non-trivial case where $n=0$. The eigenfunction with these quantum numbers is \cite{Sfetsos:1994vz}
\begin{align}
\Psi_{0,0,\delta}(\psi)\, =\, \frac{\sin \left(\frac{\delta}{k}\ell\,\psi\right)}{\sin \psi}
\end{align}
and the modular matrix for the coset is the product of modular matrices for the component WZW models (throughout this paper we will leave implied the weights $\lambda$ for the $SU(2)_k$ part, since it is not directly involved in the limit)
\begin{align}
\mathcal{S}^{\rm coset}_{(\mu,\nu),(\mu',\nu')}\, =\, \mathcal{S}^{(\ell)}_{\mu,\mu'}\, \mathcal{S}^{(k+\ell)}_{\nu,\nu'}\rightarrow \frac{2}{\ell}\sin^2 \left( \frac{\delta\delta'}{k^2}\ell\,\pi \right)
\end{align}

Using these we can do the sum (which becomes an integral at large $\ell$) corresponding to the coset Cardy state. We are interested in the behaviour of the brane on the second sphere $S^3_{\ell}$, since that is where the scaling takes place. Let us focus on its behaviour as a function of the $S^3_{\ell}$ polar coordinate $\beta_0\equiv\cos\theta$. Taking into account the normalizations in \cite{Fuchs:2005gp} we find
\begin{align}\label{eq:cosetbrane}
\mathcal{B}_{\delta} (\theta)\, =\, \frac{\ell}{k}\sqrt{\frac{2}{\ell}}\int_0^k \dd x\, \frac{\sin^2 \left( \frac{\delta\, \ell}{k^2}\,\pi \, x\right)}{\sin {\frac{\pi\, x}{k}}}\, \frac{\sin \left(\frac{\ell}{k}\, x\,\theta\right)}{\sin \theta}\,\propto\, \frac{1}{\sin\theta} H\left(\frac{2\delta}{k}\pi-\theta\right)
\end{align}
where $H(\theta)$ is the Heaviside step function: $H(\theta\geq 0)=1$, $H(\theta<0)=0$. We evaluated the integral in the large-$\ell$ limit by complex contour. This confirms that the Cardy brane is a polar cap extending down to $\theta_{\max}=(2\delta/k)\pi$ for $\delta<k/2$. We do not know how to see the shrinking for $\delta>k/2$ from this viewpoint - the answer could be to use the field identification in the coset model, which relates $\delta$ to $k-\delta$. 
\section{Branes in the NATD geometry}
The non-Abelian T-dual geometry of the $SU(2)_k$ WZW model is obtained as a limit of the coset geometry which breaks the symmetry between the two spheres. We take $\ell\rightarrow\infty$, so that the radius of $S^3_{\ell}$ becomes large. At the same time we zoom in to the North pole $\beta_0 = 1$ by a factor $1/\ell$. Explicitly, we make the coordinate transformation
\begin{align}\label{eq:natdzoom}
\alpha_0\equiv\cos\psi&&\beta_0^2\equiv 1-\left(\frac{k}{\ell}\right)^2(x^2+y^2)&&\gamma\equiv\frac{k}{\ell}\, x \sin\psi&&\ell\rightarrow\infty\quad .
\end{align}
The metric for the limiting space was written in \cite{Polychronakos:2010fg}, but the exact form will not be important for our purposes. Our central observation is that after we zoom in, the distinction between the Cardy branes seems to disappear. The `edges' of the branes, which seem to be their only defining feature at large $\ell$, are sent to infinity in the $xy$ plane. This leads to a puzzle. As show in \cite{Polychronakos:2010fg}, the modes \eqref{eq:quantumnumbers} (a continuous range of $0<\delta<k$) which are associated with these branes are certainly supported by the limiting geometry and are distinct, but there seems to be only one Cardy brane. A possible way out is to notice that in performing the limit in the integral \eqref{eq:cosetbrane} we have effectively averaged over angles of order $\Delta\theta\sim 1/\ell$, but the limit \eqref{eq:natdzoom} zooms in to a region of precisely this size. However we have investigated the integral in the limit and found no distinguishing features in the brane distributions for different $\delta$.  
\section{Fusion rules and open string spectra}
In the WZW model in section \ref{sec:wzw}, we found that zooming in to the North pole at large $k$ retained all of the Cardy branes for the corresponding closed string modes we kept. However, in the coset at large $\ell$, there were closed string modes which appeared in the spectrum of the zoomed geometry, but the corresponding Cardy branes coalesced. In this section we resolve the problem by considering the open string spectra between the Cardy branes in each case. 

A standard result is that the multiplicities of chiral algebra representations in the open string spectrum between two Cardy branes is simply given by the fusion coefficients. We consider each theory in turn. 
\subsection{WZW model}\label{sec:wzwfusion}
The fusion rules are given in terms of the modular matrices by the Verlinde formula \cite{Verlinde:1988sn}. This involves a sum over all representations of the chiral algebra. In this sum we want to consider a truncation to the  $\lambda\sim\mathcal{O}(\sqrt{k})$ weights of the WZW model, as considered in \cite{Fraser:2018cxn}. We take two scaled weights $\lambda_{1,2}=\tilde{\lambda}_{1,2}\sqrt{k}$ and consider their fusion with a third weight $\lambda_3$
\begin{align}
\mathcal{N}_{\tilde{\lambda_1}\tilde{\lambda_2}}^{(k)\ \lambda_3}|_{k\rightarrow\infty}&\propto \frac{2}{\sqrt{k}}\int_0^{x_{\rm max}}\dd x\,\frac{\sin (\pi\tilde{\lambda_1} x)\sin (\pi\tilde{\lambda_2} x)\sin (\pi\lambda_3 x)}{\sin (\frac{\pi x}{\sqrt{k}})}\\
&=\frac{1}{2}\begin{cases}
1&\textrm{if } |\tilde{\lambda}_1-\tilde{\lambda}_2|\leq\tilde{\lambda}_3\leq \tilde{\lambda}_1+\tilde{\lambda}_2\ \left(\lambda_3=\tilde{\lambda}_3\sqrt{k}\sim \sqrt{k}\right)\\
0&\textrm{otherwise}
\end{cases}
\label{eq:wzwfusion}
\end{align}
where we take $x_{\rm max}\rightarrow\infty$ after we take $k\rightarrow\infty$, in order to sum over only $\lambda\sim \sqrt{k}$. So in order to appear in the open string spectrum, $\lambda_3$ needs to scale with $\sqrt{k}$ like the other weights, and be in the interval \eqref{eq:wzwfusion}. Thus the $\sqrt{k}$ scaled Cardy brane spectrum contains only $\sqrt{k}$ scaled states. 

Geometrically, we have the following picture. The scaled region around the North pole has order one size, in units of string length. It supports closed string modes with wavelength order one. The corresponding Cardy branes are at finite radius in the scaled geometry, and open strings between any two contain only order one wavelengths. Thus we have a consistent picture of both closed and open string states. 
\subsection{Coset model}\label{sec:cosetfusion}
The analogous situation in the coset is the following. We take the level $\ell$ to infinity. We take the quantum numbers $\mu$ and $\nu$ to go linearly with $\ell$. The generic weight has $\mu-\nu=\ell$, but we only want to keep the `truncated' weights with $\mu-\nu=\mathrm{finite}$, as in \eqref{eq:quantumnumbers}. We parametrize this as
\begin{align}
&\textrm{Generic weight}:&\mu&=\frac{\tilde{\mu}}{k}\ell\, +\, \mathrm{finite}& \nu&=\frac{\tilde{\nu}}{k}\ell\, +\, \mathrm{finite}\nonumber\\
&\textrm{Truncated weight}:&\mu&=\frac{\delta}{k}\ell\, +\, \mathrm{finite}& \nu&=\frac{\delta}{k}\ell\, +\, \mathrm{finite}
\end{align}
i.e. for the truncated weights, $\tilde{\mu}=\tilde{\nu}$. 

Working to leading order in $\ell$, we now calculate the fusion coefficients between two truncated and one generic weight, as in \cite{Fraser:2018cxn}:
\begin{align}
\mathcal{N}_{\delta_1, \delta_2}^{\phantom{\delta_1, \delta_2} (\mu_3,\nu_3)}\big|_{\ell\rightarrow\infty}^{\mathrm{trunc.}}& \propto 
\frac{1}{\ell}\int_0^k\dd x\, \frac{\sin^2 \left(\pi\delta_1\ell\, \frac{x}{k}\right)\, \sin^2 \left(\pi\delta_2\ell\, \frac{x}{k}\right)}{\sin^2 \left(\pi \frac{x}{k}\right)}\sin \left(\pi\tilde{\mu}_3\ell\, \frac{x}{k}\right) \sin \left(\pi\tilde{\nu}_3\ell\, \frac{x}{k}\right)
\nonumber\\
=\, \frac{1}{16}&\sum s_2 s_3|\delta_1 + s_1 \delta_2+\frac{1}{2}(s_2\tilde{\mu}_3+s_3\tilde{\nu}_3)|\nonumber\\
-\frac{1}{8}&\sum s_1 s_2|\delta_1+\frac{1}{2}(s_1\tilde{\mu}_3+s_2\tilde{\nu}_3)|-\frac{1}{8}\sum s_1 s_2|\delta_2+\frac{1}{2}(s_1\tilde{\mu}_3+s_2\tilde{\nu}_3)|\nonumber\\
+\frac{1}{4}&\sum s_1|\frac{1}{2}(\tilde{\mu}_3+s_1\tilde{\nu}_3)|\quad .
\label{eq:cosetfusion}
\end{align}
This does not vanish when $\tilde{\mu}_3\neq\tilde{\nu}_3$. Thus the fusion rules do not close on the truncated weights. Therefore there should be no consistent closed string theory of these weights. From the open string perspective, the spectrum between two truncated Cardy branes contains generic weights. 

The geometrical picture is the following. The truncated closed string modes can be supported by the zoomed geometry \eqref{eq:natdzoom}. However, a consistent closed string theory would necessary involve other modes as well. In the zoomed geometry all of the truncated Cardy branes look the same, but this is because the open string spectrum between two of them contains extra modes which are only supported in the full coset space. 

\begin{figure}[thbp]
\centering
\begin{subfigure}[t]{0.45\textwidth}
\flushleft
\includegraphics[scale=0.7]{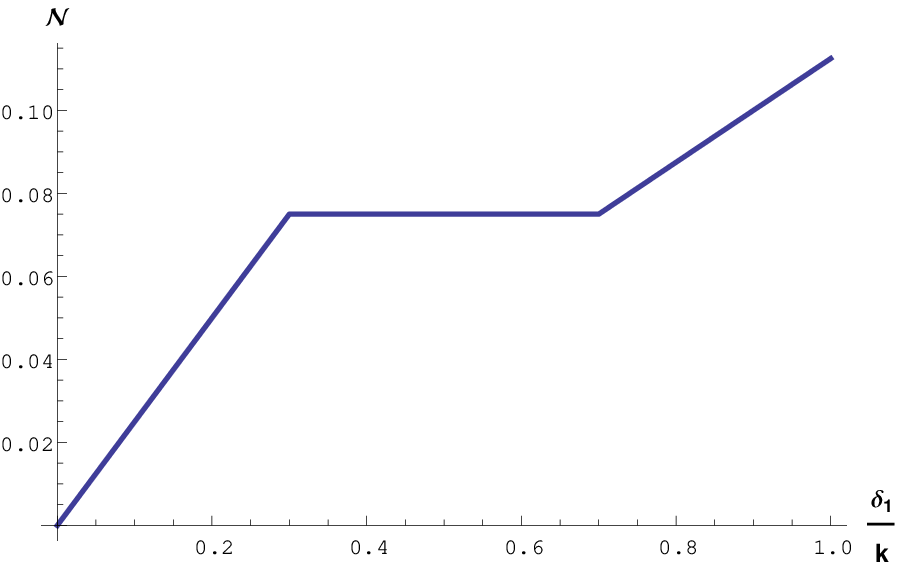}
\caption{Truncated fusion rules for modes \ $\delta_3=k$ between two Cardy branes $\delta_1$ and $\delta_2=0.3k$. }
\label{fig:fusion_uv}
\end{subfigure}
\begin{subfigure}[t]{0.45\textwidth}
\flushright
\includegraphics[scale=0.7]{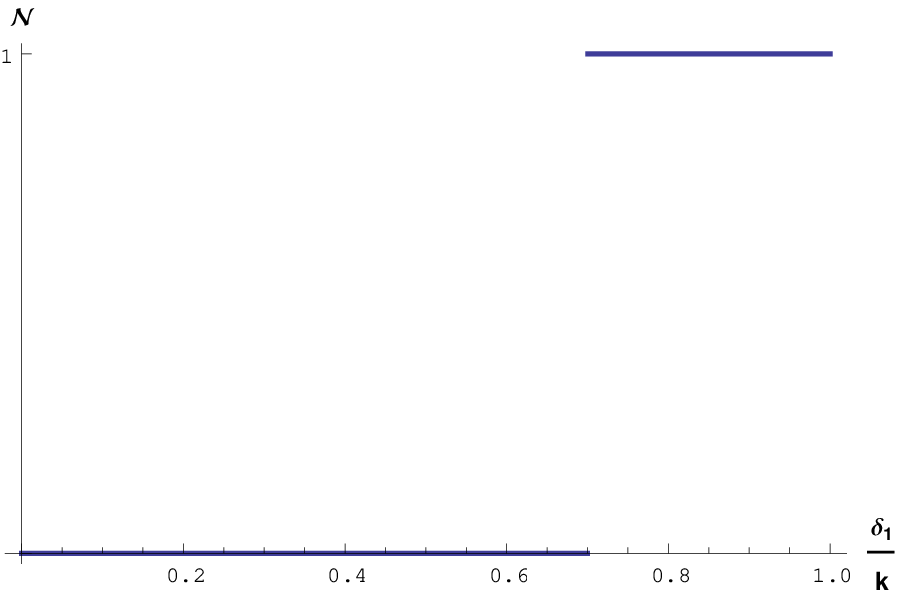}
\caption{The full fusion rules in the same case. }
\label{fig:fusion_generic}
\end{subfigure}
\caption{Coset fusion rules in the $\ell\rightarrow\infty$ limit, in the truncation vs. in the full theory. Notice the different scales: the full fusion rules are always one or zero. }
\end{figure}

The (inconsistent) truncated fusion rules \eqref{eq:cosetfusion} in fact already give a kind of geometrical picture of the the Cardy branes. Consider the case $\tilde{\mu}_3=\tilde{\nu}_3=k$, plotted in fig \ref{fig:fusion_uv}. This gives the multiplicity of these modes contained in the open string spectrum between the $\delta_1$ and $\delta_2$ branes, and for $\delta_1\leq k/2$ it is proportional to the area of overlap between the two branes. We can see them `slipping' over one another. 

However the above argument suggests we need to consider the full fusion rules where we sum over all weights, involving a double sum over $\tilde{\mu}$ and $\tilde{\nu}$ independently:
\begin{align}
&\mathcal{N}_{\delta_1, \delta_2}^{\phantom{\delta_1, \delta_2} (\mu_3,\nu_3)}\big|_{\ell\rightarrow\infty}^{\mathrm{full.}}\nonumber\\
& \propto 2\int_0^k\dd x\dd y\, \frac{\sin \left(\pi\delta_1\ell\, \frac{x}{k}\right)\sin \left(\pi\delta_1\ell\, \frac{y}{k}\right)\, \sin \left(\pi\delta_2\ell\, \frac{x}{k}\right)\sin \left(\pi\delta_2\ell\, \frac{y}{k}\right)}{\sin \left(\pi \frac{x}{k}\right)\sin \left(\pi \frac{y}{k}\right)}\nonumber\\&\hspace{3in}\times\sin \left(\pi\tilde{\mu}_3\ell\, \frac{x}{k}\right) \sin \left(\pi\tilde{\nu}_3\ell\, \frac{y}{k}\right)
\nonumber\\
&=\begin{cases}
1&\textrm{if } |\tilde{\delta}_1-\tilde{\delta}_2|\leq(\tilde{\mu}_3,\tilde{\nu}_3)\leq \tilde{\delta}_1+\tilde{\delta}_2\\
0&\textrm{otherwise}\quad .
\end{cases}
\label{eq:fullfusion}
\end{align}
This is qualitatively different from \eqref{eq:cosetfusion}. An example is depicted in fig. \ref{fig:fusion_generic}. The extra modes ($\tilde{\mu}\neq\tilde{\nu}$) that we have added do not appear in the scalar spectrum of \cite{Polychronakos:2010fg,Polychronakos:2010hd}, therefore they are stringy modes, and because of this we do not think they should affect the calculation \eqref{eq:cosetbrane}, i.e. the shape of the Cardy branes, which was based on zero modes. But adding them completely changes the fusion rules - however the geometrical interpretation of \eqref{eq:fullfusion} is less clear than \eqref{eq:cosetfusion}. 
\section{Discussion}
We argued that the geometric limit \eqref{eq:natdzoom} together with the high-spin truncation \eqref{eq:quantumnumbers} of the diagonal coset model \eqref{eq:coset} is not consistent, from the perspective of the open string picture and the related fusion rules. The distinguishing features of the different Cardy branes fall outside the zoomed-in North pole region of the geometry, and the fusion rules do not close on the truncation. 

We stress that in this note the computations were not rigorous - we were more concerned with examining the geometrical picture. In particular we always worked at leading order in $\ell$, and neglected sums over the finite weights of the algebra. It is important to fill in these details and make the arguments completely sound. The complete theory (keeping all weights) in the large $\ell$ limit should be the analogue of the $c\rightarrow\infty$ of the Virasoro minimals models, the Runkel-Watts theory \cite{Runkel:2001ng}, which corresponds to the case $k=1$ in the coset. 

From \cite{Sfetsos:1994vz} it is natural to state that the $\ell\rightarrow\infty$ limit of the coset is the non-Abelian T-dual of the $SU(2)_k$ WZW model with respect to the vector $SU(2)$ isometry. Our results suggest that if such a limit exists then it cannot have a geometric interpretation only on the scaled geometry, which is the background typically called the `non-Abelian dual'. This is further evidence to suggest that non-Abelian duals need a global completion. 
\section*{Acknowledgments}
We would like to thank Andy O'Bannon for the suggestion that led to this study, and Dimitris Manolopoulos and Kostas Sfetsos for collaboration on a closely related paper. This work is supported by the `CUniverse' research promotion project by Chulalongkorn University (grant reference CUAASC). 
\bibliographystyle{utphys}
\bibliography{branes.bib}
\end{document}